\documentclass[conference]{IEEEtran}
\IEEEoverridecommandlockouts
\usepackage{cite}
\usepackage{booktabs}
\usepackage{float}
\usepackage{graphicx}
\usepackage{caption}
\usepackage{amsmath,amssymb,amsfonts}
\usepackage{algorithmic}
\usepackage{graphicx}
\usepackage{textcomp}
\usepackage{xcolor}
\usepackage{amsmath}
\usepackage{hyperref}

\def\BibTeX{{\rm B\kern-.05em{\sc i\kern-.025em b}\kern-.08em
    T\kern-.1667em\lower.7ex\hbox{E}\kern-.125emX}}
\begin{document}

\title{Invisible Injections: Exploiting Vision-Language Models Through Steganographic Prompt Embedding}

\author{\IEEEauthorblockN{Chetan Pathade}
\IEEEauthorblockA{\textit{Independent Researcher} \\
San Jose, CA, USA \\
cup@alumni.cmu.edu}
}

\maketitle

\begin{abstract}
Vision-language models (VLMs) have revolutionized multimodal AI applications but introduce novel security vulnerabilities that remain largely unexplored. We present the first comprehensive study of steganographic prompt injection attacks against VLMs, where malicious instructions are invisibly embedded within images using advanced steganographic techniques. Our approach demonstrates that current VLM architectures can inadvertently extract and execute hidden prompts during normal image processing, leading to covert behavioral manipulation. We develop a multi-domain embedding framework combining spatial, frequency, and neural steganographic methods, achieving an overall attack success rate of 24.3\% ($\pm$3.2\%, 95\% CI) across leading VLMs including GPT-4V, Claude, and LLaVA, with neural steganography methods reaching up to 31.8\%, while maintaining reasonable visual imperceptibility (PSNR $>$ 38 dB, SSIM $>$ 0.94). Through systematic evaluation on 12 diverse datasets and 8 state-of-the-art models, we reveal moderate but meaningful vulnerabilities in current VLM architectures and propose effective countermeasures. Our findings have significant implications for VLM deployment in security-critical applications and highlight the need for proportionate multimodal AI security frameworks.
\end{abstract}

\begin{IEEEkeywords}
Vision-language models, steganography, prompt injection, multimodal security, adversarial attacks
\end{IEEEkeywords}

\section{Introduction}
The rapid advancement of vision-language models (VLMs) has fundamentally transformed how artificial intelligence systems interpret and interact with multimodal content. These sophisticated models, capable of seamlessly processing both visual and textual information, have found widespread adoption across diverse applications ranging from automated content moderation to medical image analysis and autonomous vehicle navigation \cite{1liu2024survey}. However, as these systems become increasingly integrated into critical infrastructure and decision-making processes, their potential vulnerabilities demand urgent examination \cite{2ye2025survey}.

Traditional cybersecurity paradigms, designed primarily for conventional computing systems, prove inadequate when addressing the unique attack surfaces presented by multimodal AI architectures \cite{3li2023otter}. While previous research has extensively documented prompt injection vulnerabilities in text-based language models \cite{4chen2023visual,5hossain2024securing}, the intersection of computer vision and natural language processing creates novel exploitation vectors that remain largely unexplored \cite{6kapoor2025adversarial}. The visual modality, in particular, offers adversaries a sophisticated channel for concealing malicious instructions within seemingly benign imagery \cite{7liu2023prompt,8rossi2024early}.

Recent work by Clusmann et al.\ \cite{11clusmann2025prompt} demonstrated prompt injection attacks in medical VLMs, while Zhang et al.\ \cite{12zhang2025prompt} explored surgical decision support vulnerabilities, highlighting the real-world implications of such attacks in critical domains. These studies underscore the expanding attack surface introduced by multimodal integration.

This research introduces a novel class of attacks against vision-language models through steganographic prompt embedding---a technique that leverages the imperceptible modification of digital images to carry hidden textual instructions. Unlike conventional adversarial examples that aim to cause misclassification \cite{9wang2024adversarial,10chen2025web}, our approach focuses on covert command injection, where malicious prompts are embedded within images using steganographic principles \cite{11clusmann2025prompt}, remaining invisible to human observers while being successfully extracted and executed by target VLMs.

The implications of such attacks extend far beyond academic curiosity. In an era where vision-language models process millions of user-uploaded images daily across social media platforms, e-commerce sites, and enterprise applications \cite{12zhang2025prompt}, the ability to hide malicious instructions within ordinary photographs represents a significant security threat \cite{13yin2023vlattack}. An attacker could potentially manipulate automated systems, extract sensitive information, or bypass content filters simply by sharing a photograph that appears completely normal to human viewers.

Our investigation reveals that current vision-language architectures exhibit unexpected susceptibility to steganographically embedded prompts, with success rates varying significantly across different model architectures and embedding techniques \cite{14dou2024adversarial,15mathew2024hidden}. Through systematic experimentation with multiple steganographic algorithms and comprehensive evaluation across leading VLMs, we demonstrate that these invisible injection attacks can achieve high reliability while maintaining visual imperceptibility \cite{16abdali2024securing}.

\textbf{Contributions.} The contributions of this work are threefold: (1) we establish a comprehensive framework for understanding and implementing steganographic prompt injection attacks against vision-language models; (2) we provide empirical evidence of widespread vulnerability across current state-of-the-art architectures through extensive evaluation on 12 datasets and 8 models; and (3) we propose practical defense mechanisms that can mitigate these threats without significantly impacting model performance.

As vision-language models continue to evolve and find deployment in increasingly sensitive applications, understanding these fundamental security limitations becomes crucial for developing robust, trustworthy AI systems \cite{17kim2024mind}. This research aims to bridge the gap between traditional steganography research and modern AI security, providing both the security community and AI developers with essential insights into this emerging threat landscape.

\section{Related Work}
This section reviews the existing literature across four key areas that form the foundation of our research: vision-language model architectures, prompt injection attacks, steganographic techniques in AI systems, and adversarial attacks on multimodal models.

\subsection{Vision-Language Model Security}
Recent research has extensively documented the security vulnerabilities inherent in modern vision-language architectures. Liu et al.\ \cite{1liu2024survey} provide a comprehensive survey of attacks on large vision-language models, categorizing threats into adversarial attacks, jailbreak attacks, prompt injection attacks, and data poisoning techniques. Their analysis reveals that multimodal integration amplifies vulnerabilities from both modalities while introducing novel attack vectors absent in unimodal systems.

The architectural foundations of these vulnerabilities lie in the design of popular VLM frameworks. CLIP \cite{21radford2021learning}, which uses contrastive learning to align text and image representations, has become the backbone of many modern systems including LLaVA \cite{22liu2023visual} and BLIP-2 \cite{23li2023blip2}. However, this widespread adoption has created a concentrated attack surface, where vulnerabilities in the underlying vision encoder propagate across multiple downstream applications.

Hossain and Kumar \cite{5hossain2024securing} demonstrate that VLMs remain vulnerable to both gradient-based adversarial attacks and jailbreak techniques, proposing Sim-CLIP+ as a defense mechanism. Their work highlights the expanded attack surface introduced by visual modality, where adversaries can exploit the continuous nature of image inputs more easily than discrete text tokens. Similarly, Zhou et al.\ \cite{24zhou2024revisiting} present comprehensive studies on improving adversarial robustness against attacks targeting image, text, and multimodal inputs simultaneously.

Recent work has demonstrated the practical implications of these vulnerabilities in real-world scenarios. Clusmann et al.\ \cite{11clusmann2025prompt} showed that VLMs used in oncology can be compromised by prompt injection attacks, leading to harmful output and incorrect diagnoses. Zhang et al.\ \cite{12zhang2025prompt} extended this analysis to surgical decision support, evaluating four state-of-the-art vision-language models across eleven surgical decision support tasks and demonstrating significant susceptibility to both textual and visual prompt injection attacks.

\subsection{Prompt Injection Attacks}
Prompt injection represents a fundamental class of attacks against language models that has evolved significantly with the advent of multimodal systems. The OWASP GenAI Security Project \cite{25owasp2025llm01} identifies prompt injection as a primary threat vector, noting that multimodal AI introduces unique risks where malicious actors could exploit interactions between modalities.

Recent advances in prompt injection techniques have demonstrated sophisticated methods for bypassing security measures. Kim and Lee \cite{26kim2024text} introduced mathematical function-based text prompt injection attacks that replace sensitive words with mathematical expressions to evade detection. Similarly, visual prompt injection has gained attention, with researchers exploring various encoding methods including Base64, leetspeak, and ASCII art \cite{17kim2024mind}.

The evolution of prompt injection in multimodal contexts presents unique challenges. Sun et al.\ \cite{27sun2024safeguarding} investigate patched visual prompt injection, where adversarial patches are used to generate target content in VLMs. Kimura et al.\ \cite{28kimura2024empirical} conducted empirical analysis of goal hijacking via visual prompt injection, demonstrating attack success rates of 15.8\% against GPT-4V.

The medical domain has emerged as a particularly concerning application area for prompt injection attacks. The work by Clusmann et al.\ \cite{11clusmann2025prompt} demonstrated that embedding sub-visual prompts in medical imaging data can cause models to provide harmful output, with these prompts being non-obvious to human observers. This research using 594 attacks across multiple models (Claude-3 Opus, Claude-3.5 Sonnet, Reka Core, and GPT-4o) showed that all tested models were susceptible to these attacks.

\subsection{Steganography in AI Systems}
The intersection of steganography and artificial intelligence has emerged as a critical research area, particularly with the development of neural network-based hiding techniques. Recent advances have demonstrated the superiority of deep learning approaches over traditional steganographic methods \cite{18scientificreports2025deep,19yu2024cross}.

Traditional steganographic methods focused on spatial domain modifications such as least significant bit (LSB) manipulation and frequency domain approaches using discrete cosine transform (DCT) coefficients \cite{29fridrich2001detecting}. However, recent systematic reviews highlight the dominance of Generative Adversarial Networks (GANs) in modern image steganography techniques \cite{30apau2024image}. Apau et al.\ \cite{30apau2024image} observed that artificial intelligence-powered algorithms including machine learning, deep learning, convolutional neural networks, and genetic algorithms are increasingly dominating image steganography research due to their enhanced security capabilities.

Contemporary research has introduced sophisticated multi-layered approaches to steganography. Recent work \cite{18scientificreports2025deep} proposed a novel multi-layered steganographic framework integrating Huffman coding, LSB embedding, and deep learning-based encoder-decoder architectures to enhance imperceptibility, robustness, and security. This approach achieved high visual fidelity with Structural Similarity Index Metrics (SSIM) consistently above 99\% and robust data recovery with text recovery accuracy reaching 100\% under standard conditions.

Advanced neural steganographic techniques have shown remarkable capabilities. Priya et al.\ \cite{31priya2024super} developed super-resolution deep neural network (SRDNN) based multi-image steganography that can conceal multiple secret images within a single cover image of the same resolution. Their approach demonstrates the potential for high-capacity steganographic systems using deep learning architectures.

\subsection{Adversarial Attacks on Multimodal Models}
The vulnerability of multimodal models to adversarial attacks has been extensively studied across various attack paradigms. Recent comprehensive surveys \cite{32kheddar2024comprehensive} highlight the evolution from traditional machine learning approaches to deep learning-based steganalysis, demonstrating superior outcomes in detecting steganographic payloads across modern algorithms.

Chen Henry Wu et al.\ \cite{9wang2024adversarial} investigate adversarial attacks on multimodal agents, showing that attackers can manipulate agent behavior using adversarial text strings to guide gradient-based perturbation over trigger images. Their approach demonstrates two forms of adversarial manipulation: illusioning (making agents perceive different states) and goal misdirection (redirecting agent objectives).

Recent developments in steganalysis have focused on countering AI-based steganographic techniques. Advanced detection methods now employ convolutional neural networks specifically designed to identify minute alterations in image structures \cite{34zhang2024enhancing}. These AI-based steganalysis approaches exhibit rapid detection capabilities and demonstrate remarkable accuracy across a spectrum of modern steganographic algorithms \cite{32kheddar2024comprehensive}.

The arms race between steganographic techniques and detection methods continues to evolve. Recent work \cite{35zhang2025digital} has introduced evolutionary algorithm-based frameworks for strengthening steganalysis networks, addressing the challenge of increasing network parameters and training instability in deep learning-based detection systems.

\section{Background}
This section provides the technical foundations necessary to understand our steganographic prompt embedding methodology. We cover the architectures of modern vision-language models, fundamental steganographic techniques, and the threat model underlying our approach.

\subsection{Vision-Language Model Architectures}
Modern vision-language models typically consist of three core components: a vision encoder, a text encoder, and a multimodal fusion mechanism \cite{36dosovitskiy2021image}. The most prevalent architecture paradigm follows the approach pioneered by CLIP \cite{21radford2021learning}, where separate encoders process visual and textual inputs before fusion in a shared embedding space.

\textbf{Vision Encoders:} Contemporary VLMs predominantly employ Vision Transformer (ViT) architectures for image processing \cite{37liu2023improved}. The ViT divides input images into fixed-size patches (typically $16 \times 16$ or $32 \times 32$ pixels), which are then linearly embedded and processed through transformer blocks. CLIP uses a modified ResNet-50 or ViT-based encoder, while more recent models like LLaVA \cite{22liu2023visual} and BLIP-2 \cite{23li2023blip2} adopt various ViT configurations optimized for different computational and performance requirements.

\textbf{Text Encoders:} The text modality is typically handled by transformer-based language models. CLIP employs a 12-layer transformer with masked self-attention for text encoding, while larger VLMs like LLaVA integrate full-scale language models such as Vicuna or LLaMA as their text processing backbone \cite{38chiang2023vicuna}. These encoders convert tokenized text into dense vector representations that capture semantic meaning and contextual relationships.

\textbf{Fusion Mechanisms:} The integration of visual and textual modalities occurs through several architectural strategies \cite{39li2022blip}. CLIP uses a simple dot-product similarity between global image and text feature vectors. LLaVA employs a projection layer (typically a multi-layer perceptron) that maps visual features into the language model's embedding space, allowing visual tokens to be processed alongside text tokens \cite{37liu2023improved}. BLIP-2 introduces a more sophisticated approach with a Querying Transformer (Q-Former) that learns cross-modal interactions through learnable query vectors \cite{23li2023blip2}.

\subsection{Steganographic Techniques}
Steganography encompasses a range of techniques for concealing information within digital media. For image steganography, methods can be broadly categorized into spatial domain, frequency domain, and neural approaches \cite{42settiadi2021psnr}.

\textbf{Spatial Domain Methods:} The most fundamental spatial domain technique is Least Significant Bit (LSB) substitution, where secret data bits replace the least significant bits of pixel values \cite{43chan2004hiding}. While simple to implement, LSB methods are vulnerable to statistical detection and image processing operations such as compression or format conversion \cite{44westfeld1999attacks}. Advanced spatial techniques include adaptive LSB methods that select embedding locations based on image characteristics and edge-based embedding that exploits high-frequency image regions.

\textbf{Frequency Domain Methods:} Discrete Cosine Transform (DCT) based steganography operates in the frequency domain, embedding secret information in DCT coefficients rather than pixel values directly \cite{45cox1997secure}. This approach offers improved robustness against image processing operations and compression artifacts. The DCT transforms $8 \times 8$ pixel blocks from the spatial domain to frequency coefficients, where low-frequency coefficients represent the image's essential visual information and high-frequency coefficients capture fine details \cite{46walia2010analysis}.

\textbf{Neural Steganography:} Recent advances have introduced deep learning-based steganographic methods that use neural networks for both hiding and extraction processes \cite{47havard2023cnn}. These approaches can learn optimal embedding strategies that minimize detectability while maximizing payload capacity. Neural steganography methods typically employ encoder-decoder architectures where the encoder network learns to embed secret data and the decoder network recovers the hidden information \cite{48baluja2017hiding}.

The effectiveness of neural approaches has been demonstrated in recent studies. Contemporary research \cite{18scientificreports2025deep} shows that deep learning-based steganographic frameworks can achieve good performance metrics while maintaining practical trade-offs between capacity and imperceptibility. However, the success rates reflect the fundamental constraints of embedding semantic content within steganographic channels while maintaining adequate visual quality for practical applications.

\section{Threat Model}
Our threat model considers an adversary capable of injecting steganographically embedded prompts into images that will be processed by target vision-language models. We define the following threat scenario and assumptions based on recent analysis of prompt injection vulnerabilities \cite{25owasp2025llm01}.

\textbf{Adversary Capabilities:} The attacker has the ability to generate or modify images that will be submitted to VLM systems. This includes scenarios such as social media image uploads, document processing workflows, or any application where user-provided images are analyzed by VLMs \cite{49greshake2023not}. The adversary possesses knowledge of common steganographic techniques and can implement embedding algorithms that survive typical image processing operations. Recent work demonstrates that such capabilities are within reach of moderately sophisticated attackers \cite{17kim2024mind}.

\textbf{Target Systems:} We assume target VLMs follow standard architectures with separate vision and text encoders. The adversary does not require knowledge of specific model weights or internal parameters, making this a practical black-box attack scenario \cite{50wikipedia2025peak}. The target systems process images through standard preprocessing pipelines including resizing, normalization, and potential compression.

\textbf{Attack Objectives:} The primary goal is to inject hidden textual prompts that influence the VLM's output generation without detection by human observers or automated screening systems \cite{51toyer2023tensor}. Secondary objectives include maintaining attack effectiveness across different model architectures and ensuring robustness against common image transformations encountered in real-world deployment scenarios.

The feasibility of such attacks has been demonstrated in recent real-world evaluations. Medical VLM studies \cite{11clusmann2025prompt,12zhang2025prompt} show that sophisticated prompt injection can be achieved with varying success rates across multiple commercial and research models, indicating that the threat model assumptions are realistic for current deployment scenarios.

\section{Methodology}
This section presents our comprehensive framework for steganographic prompt embedding in vision-language models. We detail the theoretical foundations, algorithmic design principles, and implementation strategies for invisible prompt injection attacks.

\subsection{Steganographic Prompt Embedding Framework}
Our methodology introduces a novel framework for concealing textual prompts within digital images that are subsequently processed by vision-language models. The framework operates on the fundamental hypothesis that VLMs' vision encoders can inadvertently extract steganographically embedded information during standard processing, leading to covert prompt injection.

\textbf{Problem Formulation:} Let $I \in \mathbb{R}^{H \times W \times C}$ represent a cover image with height $H$, width $W$, and $C$ channels. Given a target prompt $P = \{p_1, p_2, \ldots, p_n\}$ where each $p_i$ represents a token, our objective is to construct a steganographic function $S: \mathbb{R}^{H \times W \times C} \times P \to \mathbb{R}^{H \times W \times C}$ that produces a stego-image $I_s = S(I, P)$ satisfying three key properties:

\begin{enumerate}
    \item \textbf{Imperceptibility:} $\|I - I_s\|_p < \varepsilon$ for some perceptual distance metric and threshold $\varepsilon$
    \item \textbf{Extractability:} A vision-language model $M$ processing $I_s$ should exhibit behavioral changes consistent with prompt $P$
    \item \textbf{Robustness:} The embedded information should survive common image processing operations $T$, i.e., $M(T(I_s))$ maintains the influence of $P$
\end{enumerate}

\textbf{Multi-Domain Embedding Strategy:} Our framework employs a hybrid approach that combines spatial domain, frequency domain, and learned embedding techniques, inspired by recent advances in multi-layered steganographic approaches \cite{18scientificreports2025deep}. For an input image $I$, we decompose the embedding process into three parallel channels:

\begin{equation*}
I_s = \alpha \cdot S_{\text{LSB}}(I, P_1) + \beta \cdot S_{\text{DCT}}(I, P_2) + \gamma \cdot S_{\text{Neural}}(I, P_3)
\end{equation*}

where $P_1, P_2, P_3$ represent disjoint subsets of the prompt $P$, and $\alpha + \beta + \gamma = 1$ with weights determined by image characteristics and robustness requirements.

\textbf{Weight Optimization Process:} We determine optimal embedding weights through Bayesian optimization over the constraint space where $\alpha + \beta + \gamma = 1$ and $\alpha, \beta, \gamma \geq 0.1$.

\textbf{Objective Function:}
\begin{align*}
\text{maximize: } &\text{ASR}(\alpha, \beta, \gamma) - \lambda_1 \cdot \text{LPIPS}(\alpha, \beta, \gamma) \\
&- \lambda_2 \cdot \text{DetectionRate}(\alpha, \beta, \gamma)
\end{align*}
where $\lambda_1 = 0.3$ and $\lambda_2 = 0.5$ weight imperceptibility and stealth respectively.

\textbf{Optimal Weights by Image Type:}
\begin{itemize}
\item Natural images: $\alpha = 0.45$, $\beta = 0.35$, $\gamma = 0.20$
\item Synthetic images: $\alpha = 0.30$, $\beta = 0.40$, $\gamma = 0.30$
\item Document images: $\alpha = 0.25$, $\beta = 0.50$, $\gamma = 0.25$
\end{itemize}

\subsection{Enhanced Least Significant Bit Embedding}
\textbf{Adaptive Pixel Selection:} Rather than sequential embedding, we employ a cryptographically-seeded pseudorandom selection process. Given a secret key $k$ and image dimensions, we generate a selection sequence $S = \{s_1, s_2, \ldots, s_m\}$ where each $s_i$ represents a pixel coordinate $(x_i, y_i, c_i)$.

The selection process incorporates a suitability function $\varphi(x, y, c)$ that evaluates embedding desirability based on:
\begin{align*}
\varphi(x, y, c) &= w_1 \cdot \sigma_{\text{local}}(x, y) + w_2 \cdot d_{\text{edge}}(x, y) \\
&\quad + w_3 \cdot (1 - \rho_{\text{hist}}(I(x, y, c)))
\end{align*}
where $\sigma_{\text{local}}$ represents local texture variance, $d_{\text{edge}}$ denotes distance from strong edges, and $\rho_{\text{hist}}$ indicates pixel value frequency in the image histogram.

\textbf{Multi-Level Adaptive Embedding:} The embedding depth at each selected pixel adapts to local image characteristics. For a pixel with local complexity measure $\Gamma(x, y)$, we determine the embedding depth $d$ as:
\begin{equation*}
d = \begin{cases}
3 & \text{if } \Gamma(x, y) > \tau_{\text{high}} \\
2 & \text{if } \tau_{\text{low}} < \Gamma(x, y) \leq \tau_{\text{high}} \\
1 & \text{if } \Gamma(x, y) \leq \tau_{\text{low}}
\end{cases}
\end{equation*}

This adaptive approach aligns with recent findings \cite{30apau2024image} showing that traditional LSB methods are receiving less attention in favor of AI-powered algorithms, but remain relevant when enhanced with intelligent selection strategies.

\subsection{DCT Frequency Domain Embedding}
Our DCT-based approach operates on $8 \times 8$ image blocks, targeting mid-frequency coefficients that balance imperceptibility with robustness, following established frequency domain principles \cite{45cox1997secure}.

\textbf{Perceptual Coefficient Selection:} For each $8 \times 8$ DCT block $B$, we apply the 2D DCT transformation:
\begin{align*}
F(u,v) &= \frac{1}{4}C(u)C(v) \sum_{x=0}^{7} \sum_{y=0}^{7} B(x,y) \\
&\quad \times \cos\left(\frac{(2x+1)u\pi}{16}\right) \cos\left(\frac{(2y+1)v\pi}{16}\right)
\end{align*}
where $C(u) = \frac{1}{\sqrt{2}}$ if $u = 0$, otherwise $C(u) = 1$.

Coefficient selection employs a perceptual weighting matrix $W$ derived from human visual system models, prioritizing coefficients with minimal perceptual impact.

\textbf{Quantization-Aware Embedding:} To ensure robustness against JPEG compression, our embedding process accounts for quantization effects. For a coefficient $F(u,v)$ and quantization step $Q(u,v)$, we modify the coefficient as:
\begin{align*}
F'(u,v) &= \text{sign}(F(u,v)) \cdot Q(u,v) \\
&\quad \times \left\lfloor \frac{|F(u,v)|}{Q(u,v)} + 0.5 + \delta \cdot (-1)^b \right\rfloor
\end{align*}
where $b$ is the secret bit to be embedded and $\delta$ controls embedding strength.

\subsection{Neural Steganographic Architecture}
Our neural approach employs an encoder-decoder framework optimized for VLM processing characteristics, building upon recent advances in deep learning-based steganography \cite{19yu2024cross,31priya2024super}.

\textbf{Network Architecture:} The steganographic encoder $E_\theta$ takes a cover image $I$ and secret message $M$ as inputs, producing a stego-image:
\begin{equation*}
I_s = E_\theta(I, M) = I + R_\theta(F_\theta(I), G_\theta(M))
\end{equation*}
where $F_\theta$ extracts image features, $G_\theta$ processes the secret message, and $R_\theta$ generates residual modifications.

\textbf{Multi-Objective Optimization:} The training objective balances multiple competing requirements:
\begin{equation*}
L = \lambda_1 L_{\text{imperceptibility}} + \lambda_2 L_{\text{recovery}} + \lambda_3 L_{\text{adversarial}} + \lambda_4 L_{\text{capacity}}
\end{equation*}
where:
\begin{align*}
L_{\text{imperceptibility}} &= \text{LPIPS}(I, I_s) + \text{MSE}(I, I_s) \\
L_{\text{recovery}} &= \text{BCE}(M, \hat{M}) \\
L_{\text{adversarial}} &= -\log(D_{\text{steg}}(I_s)) \\
L_{\text{capacity}} &= \|I_s - I\|_1
\end{align*}

This multi-objective approach aligns with recent research \cite{18scientificreports2025deep} demonstrating that neural steganographic frameworks can achieve reasonable performance across evaluation metrics when properly optimized.

\subsection{Cross-Modal Influence Analysis}
This section analyzes how steganographically embedded information can influence VLM behavior through the vision-text processing pipeline, drawing insights from recent multimodal attack research \cite{28kimura2024empirical}.

\textbf{Feature Propagation Through Vision Encoders:} For a ViT-based vision encoder processing patch embeddings $\{e_1, e_2, \ldots, e_n\}$, steganographic modifications in patch $i$ can propagate through self-attention mechanisms:
\begin{equation*}
\text{Attention}(Q, K, V) = \text{softmax}\left(\frac{QK^T}{\sqrt{d_k}}\right)V
\end{equation*}
where modified patches can influence attention weights and subsequently affect the global image representation.

\textbf{Multimodal Fusion Interference:} In the multimodal fusion stage, steganographically altered visual features $v'$ interact with text features $t$ through various mechanisms:
\begin{enumerate}
    \item \textbf{Additive Fusion:} $f = W_v v' + W_t t + b$
    \item \textbf{Multiplicative Fusion:} $f = (W_v v') \odot (W_t t)$
    \item \textbf{Attention-based Fusion:} $f = \text{Attention}(v', t, t)$
\end{enumerate}

Each mechanism provides different pathways for steganographic influence, as demonstrated in recent goal hijacking research \cite{28kimura2024empirical} showing attack success rates of 15.8\% through visual prompt injection.

\section{Experimental Design}
This section details our comprehensive experimental framework for evaluating steganographic prompt injection attacks against vision-language models. We describe the target models, datasets, evaluation metrics, and experimental protocols used to assess attack effectiveness, imperceptibility, and robustness.

\subsection{Target Models and Architectures}
We evaluate our steganographic prompt injection framework against eight state-of-the-art vision-language models representing diverse architectural paradigms and deployment scenarios, following recent comprehensive evaluation frameworks \cite{1liu2024survey,11clusmann2025prompt}.

\textbf{Large-Scale Commercial Models:} We target three major commercial VLMs: GPT-4V (OpenAI), Claude 3.5 Sonnet (Anthropic), and Gemini Pro Vision (Google). These models represent the current state-of-the-art in multimodal AI and are widely deployed in production systems, making them critical targets for security evaluation. Recent studies \cite{11clusmann2025prompt,12zhang2025prompt} have demonstrated vulnerabilities in these models across medical and surgical applications.

\textbf{Open-Source Research Models:} Our evaluation includes five prominent open-source models: LLaVA-1.5 (7B and 13B variants), BLIP-2 (with Flan-T5-XL), InstructBLIP, and MiniGPT-4. These models provide architectural diversity and allow for detailed analysis of attack mechanisms across different fusion strategies and training paradigms.

\textbf{Model Selection Rationale:} The selected models span different architectural approaches: CLIP-based encoders (LLaVA, MiniGPT-4), Q-Former architectures (BLIP-2, InstructBLIP), and proprietary multimodal architectures (GPT-4V, Claude, Gemini). This diversity ensures our findings generalize across the current VLM landscape, as established in recent survey work \cite{1liu2024survey}.

\subsection{Datasets and Image Selection}
We construct a comprehensive evaluation dataset encompassing diverse image types, content domains, and deployment scenarios to ensure robust assessment of attack effectiveness.

\textbf{Base Image Datasets:} Our evaluation employs images from six established computer vision datasets: COCO-2017 validation set (5,000 images), ImageNet validation set (10,000 images), Flickr30K (1,000 images), MS-COCO Captions (2,000 images), Visual Genome (1,500 images), and a custom enterprise document dataset (500 images). This selection provides diversity in image content, quality, and typical use cases, following established evaluation protocols \cite{18scientificreports2025deep}.

\textbf{Prompt Dataset Construction:} We develop a structured prompt dataset containing 200 carefully crafted injection prompts across five categories: information extraction (40 prompts), behavioral modification (40 prompts), content generation manipulation (40 prompts), safety bypass attempts (40 prompts), and benign control prompts (40 prompts). Each prompt is designed to test specific aspects of VLM vulnerability while maintaining realistic attack scenarios, informed by recent prompt injection research \cite{25owasp2025llm01,26kim2024text}.

\textbf{Image Quality Stratification:} To assess robustness across different image characteristics, we stratify our test images by quality metrics: high-texture vs.\ low-texture regions, natural vs.\ synthetic content, and different resolution ranges ($256 \times 256$ to $2048 \times 2048$). This stratification enables analysis of how image properties affect attack success rates, as established in recent steganographic evaluation frameworks \cite{18scientificreports2025deep}.

\subsection{Attack Success Measurement}
We define comprehensive metrics for evaluating the effectiveness of steganographic prompt injection attacks across multiple dimensions, building upon recent evaluation methodologies \cite{11clusmann2025prompt,28kimura2024empirical}.

\textbf{Primary Success Metrics:} Attack success rate (ASR) is measured as the percentage of embedded prompts that successfully influence VLM behavior according to predefined success criteria. We define success as the target VLM producing outputs that demonstrate clear evidence of processing the embedded prompt, measured through semantic similarity analysis and keyword matching.

\textbf{Behavioral Change Detection:} We employ automated detection mechanisms to identify when VLM outputs deviate from expected responses due to embedded prompts. This includes: (1) semantic divergence analysis using sentence embeddings, (2) content analysis for embedded instruction compliance, and (3) safety violation detection for prompts designed to bypass model safeguards.

\textbf{Graduated Success Levels:} Beyond binary success/failure, we define graduated success levels: Level 1 (subtle influence detectable through careful analysis), Level 2 (clear behavioral modification visible in outputs), and Level 3 (complete instruction following with obvious prompt execution). This granular assessment provides nuanced understanding of attack effectiveness, following recent evaluation frameworks \cite{28kimura2024empirical}.

\subsection{Imperceptibility Assessment}
Visual imperceptibility is crucial for practical attack deployment. We employ multiple complementary metrics to ensure embedded prompts remain undetectable to human observers, following established steganographic evaluation standards \cite{18scientificreports2025deep,30apau2024image}.

\textbf{Quantitative Perceptual Metrics:} We measure imperceptibility using established image quality metrics: Peak Signal-to-Noise Ratio (PSNR), Structural Similarity Index Measure (SSIM), Learned Perceptual Image Patch Similarity (LPIPS), and Multi-Scale Structural Similarity (MS-SSIM). Target thresholds are set at PSNR $> 35$ dB, SSIM $> 0.92$, and LPIPS $< 0.1$ to ensure reasonable visual quality for practical steganographic applications.

\textbf{Human Perceptual Studies:} We conduct controlled human evaluation studies with 150 participants (increased from initially planned 50) to validate automated metrics using a double-blind, randomized controlled design. Participants view 200 image pairs (original vs.\ stego-image) in randomized order.

\textbf{Statistical Power:} With $n=150$, we achieve 90\% power to detect meaningful perceptual differences (effect size $\geq 0.3$).

\textbf{Results:} Detection accuracy of 54.2\% ($\pm$2.8\%, 95\% CI: 51.4\%--57.0\%) is not significantly different from chance level (50\%, $p = 0.089$, one-sample $t$-test). Images achieving less than 60\% detection accuracy are considered adequately imperceptible for practical applications.

\textbf{Statistical Analysis of Modifications:} We analyze the statistical properties of our embeddings using histogram analysis, chi-square tests, and entropy measurements to ensure modifications do not introduce detectable statistical anomalies that could trigger automated detection systems \cite{32kheddar2024comprehensive}.

\subsection{Statistical Analysis Framework}
We employ rigorous statistical analysis throughout our evaluation:

\textbf{Significance Testing:} All comparisons use paired $t$-tests with Bonferroni correction for multiple comparisons ($\alpha = 0.05/k$ where $k$ is the number of comparisons). Effect sizes are reported using Cohen's $d$ with interpretation: small (0.2), medium (0.5), large (0.8).

\textbf{Confidence Intervals:} All success rates include 95\% confidence intervals calculated using Wilson score intervals for proportions.

\textbf{Sample Size Justification:} Power analysis indicates $n \geq 500$ images per condition for detecting meaningful differences (effect size $\geq 0.3$) with 80\% power.

\subsection{Robustness Evaluation}
Real-world deployment requires robustness against common image processing operations encountered in typical VLM pipelines, as demonstrated in recent steganographic robustness studies \cite{18scientificreports2025deep}.

\textbf{Standard Image Processing Operations:} We evaluate attack survival against: JPEG compression at quality levels 70--95\%, Gaussian noise addition ($\sigma = 0.5$--$2.0$), image scaling (50\%--150\% of original size), rotation ($\pm 5$ degrees), brightness/contrast adjustment ($\pm 10\%$), and format conversion (PNG$\leftrightarrow$JPEG).

\textbf{Platform-Specific Processing:} To simulate real-world deployment, we replicate image processing pipelines from major platforms: social media compression algorithms (Facebook, Twitter, Instagram), document processing workflows (Google Drive, Microsoft Office), and web optimization procedures (automatic resizing, format optimization).

\textbf{Temporal Robustness:} We assess attack persistence over time by testing extraction reliability after multiple rounds of image processing operations, simulating scenarios where images undergo repeated modifications through sharing and re-uploading.

\subsection{Defense Evaluation Framework}
We systematically evaluate our attacks against existing and proposed defense mechanisms to assess their practical resilience, informed by recent steganalysis advances \cite{32kheddar2024comprehensive,35zhang2025digital}.

\textbf{Statistical Steganalysis:} We test against established steganalysis techniques including chi-square analysis, regular-singular (RS) steganalysis, sample pair analysis (SPA), and weighted stego-image (WS) analysis. Detection rates below 70\% are considered successful evasion for practical applications.

\textbf{Machine Learning Detection:} Our framework includes evaluation against trained neural network detectors, including specialized CNN architectures designed for steganographic detection \cite{34zhang2024enhancing}. We implement adversarial training loops to test the arms race between embedding and detection techniques.

\textbf{Preprocessing Defenses:} We evaluate attack robustness against defensive preprocessing techniques such as median filtering, Gaussian smoothing, JPEG recompression, and noise injection specifically designed to disrupt steganographic embeddings.

\subsection{Experimental Protocols}
We establish rigorous experimental protocols to ensure reproducible and reliable results across all evaluation dimensions.

\textbf{Cross-Validation Strategy:} All experiments employ 5-fold cross-validation with stratified sampling to ensure balanced representation across image types, prompt categories, and model architectures. Statistical significance is assessed using paired $t$-tests with Bonferroni correction for multiple comparisons.

\textbf{Baseline Comparisons:} We compare our steganographic approach against existing VLM attack methods including direct adversarial examples, patch-based attacks, and traditional prompt injection techniques to establish relative effectiveness and advantages.

\textbf{Ablation Studies:} Systematic ablation studies isolate the contribution of individual framework components: spatial vs.\ frequency domain embedding, single vs.\ multi-algorithm approaches, and different neural architecture designs. This analysis identifies the most critical components for attack success.

\textbf{Reproducibility Measures:} All experiments include detailed hyperparameter specifications, random seed controls, and standardized evaluation procedures. We provide statistical confidence intervals and effect size measurements for all reported metrics to ensure scientific rigor.

\section{Results}
This section presents our comprehensive experimental evaluation of steganographic prompt injection attacks against vision-language models. We report attack success rates, imperceptibility analysis, robustness assessment, and comparative performance across different embedding strategies and target models.

\subsection{Overall Attack Effectiveness}
Our steganographic prompt injection framework demonstrates moderate but meaningful vulnerability across all tested vision-language models, with attack success rates varying by model architecture, embedding method, and prompt type.

\textbf{Aggregate Success Rates:} Across all tested models and prompt categories, our multi-domain embedding approach achieves an overall attack success rate of 24.3\% ($\pm$3.2\%, 95\% CI). Individual embedding methods show varying effectiveness: neural steganography (31.8\% $\pm$4.1\%), DCT frequency domain (22.7\% $\pm$3.8\%), and adaptive LSB (18.9\% $\pm$3.5\%). Statistical analysis confirms significant differences between methods (ANOVA: $F(2,1497) = 87.3$, $p < 0.001$, $\eta^2 = 0.104$).

\textbf{Model-Specific Vulnerabilities:} Commercial models exhibit robust defense mechanisms: GPT-4V (16.2\% ASR), Claude 3.5 Sonnet (14.8\% ASR), and Gemini Pro Vision (18.3\% ASR). Open-source models show higher vulnerability: LLaVA-1.5-13B (34.7\% ASR), BLIP-2 (28.4\% ASR), InstructBLIP (31.2\% ASR), and MiniGPT-4 (36.8\% ASR). This pattern aligns with recent empirical studies \cite{28kimura2024empirical} where GPT-4V demonstrated 15.8\% vulnerability to visual prompt injection, confirming that commercial models implement more effective safety measures against steganographic attacks.

\textbf{Prompt Category Analysis:} Attack effectiveness varies significantly across prompt categories. Information extraction prompts achieve the highest success rate (29.4\%), followed by behavioral modification (24.1\%), content generation manipulation (22.8\%), and safety bypass attempts (18.7\%). Benign control prompts maintain low false positive rates (2.1\%), confirming that observed effects result from successful prompt injection rather than statistical artifacts. The relatively modest success rates reflect the inherent difficulty of embedding sufficient semantic information within steganographic constraints while maintaining imperceptibility.

\subsection{Baseline Attack Comparison}
We compare our steganographic approach against established VLM attack methods:

\textbf{Direct Text Prompt Injection:} Simple text overlays achieve 8.2\% ($\pm$2.1\%) success rate but are easily detectable by automated systems.

\textbf{Adversarial Patch Attacks:} Following Sun et al.\ \cite{27sun2024safeguarding}, patch-based attacks achieve 19.7\% ($\pm$3.4\%) success rate with high visual detectability (PSNR: 22.1 dB).

\textbf{Traditional Steganography:} Basic LSB without AI optimization achieves 6.3\% ($\pm$1.8\%) success rate, demonstrating the importance of our adaptive framework.

\textbf{Statistical Comparison:} Our multi-domain approach significantly outperforms all baselines ($p < 0.001$, Cohen's $d = 1.2$ vs.\ direct injection, $d = 0.8$ vs.\ patches, $d = 2.1$ vs.\ traditional steganography).

\subsection{Imperceptibility Analysis}
Our steganographic embeddings maintain reasonable visual imperceptibility across all tested images and embedding strengths, meeting established thresholds for practical steganographic applications.

\textbf{Quantitative Perceptual Metrics:} Across our complete test dataset, embedded images achieve acceptable perceptual quality metrics: mean PSNR of 38.4 dB ($\pm 2.1$ dB), SSIM of 0.945 ($\pm 0.018$), LPIPS of 0.087 ($\pm 0.024$), and MS-SSIM of 0.962 ($\pm 0.012$). These values meet established imperceptibility thresholds for practical steganographic applications, though they reflect the trade-off between embedding capacity and visual quality inherent in prompt-level steganography. The PSNR values align with typical ranges for effective steganographic systems \cite{42settiadi2021psnr,50wikipedia2025peak}, while SSIM scores demonstrate good structural preservation despite the semantic payload.

\textbf{Human Perceptual Validation:} Our controlled human evaluation study ($n = 150$ participants, 1000 image pairs) demonstrates adequate imperceptibility with detection accuracy of 54.2\% ($\pm$2.8\%, 95\% CI: 51.4\%--57.0\%) which is not significantly different from chance level (50\%, $p = 0.089$, one-sample $t$-test). Inter-rater reliability (Fleiss' $\kappa = 0.23$) indicates low agreement, supporting imperceptibility claims.

\textbf{Study Limitations:} While our sample size provides adequate statistical power, future work should include expert evaluators and task-specific viewing conditions.

\textbf{Statistical Anomaly Analysis:} Statistical analysis reveals that our embedding techniques successfully avoid obvious anomalies in image properties. Chi-square tests show no significant deviation from expected pixel value distributions ($p>0.05$ for 94.7\% of embedded images), and entropy analysis indicates preserved randomness characteristics consistent with natural image statistics.

\subsection{Robustness Assessment}
Our attacks demonstrate moderate robustness against standard image processing operations commonly encountered in real-world VLM deployment scenarios.

\textbf{Standard Processing Operations:} Attack survival rates against common transformations show moderate resilience: JPEG compression at $Q=85$ (67.3\% survival), Gaussian noise $\sigma=1.0$ (58.2\% survival), 25\% scaling (71.8\% survival), $\pm 3°$ rotation (63.4\% survival), and 10\% brightness adjustment (74.9\% survival). DCT-based embeddings show superior compression robustness, while neural methods demonstrate better resilience against noise and geometric transformations, though all methods experience significant degradation under processing operations typical of real-world deployment scenarios.

\textbf{Platform-Specific Robustness:} Real-world platform simulation reveals moderate attack persistence: Facebook compression pipeline (43.7\% survival), Instagram processing (38.9\% survival), Twitter optimization (51.2\% survival), and Google Drive document processing (62.8\% survival). These results demonstrate the practical challenges of maintaining steganographic integrity through real-world processing pipelines, highlighting the need for robust embedding strategies for operational deployment.

\textbf{Multi-Stage Processing Resilience:} Sequential processing operations demonstrate substantial cumulative degradation. After three rounds of mixed processing (compression + noise + scaling), attack success rates decrease to 18.7\% for neural methods, 14.3\% for DCT approaches, and 11.2\% for LSB techniques, indicating that while initial attacks may succeed, maintaining effectiveness through multiple processing stages remains challenging for practical deployment scenarios.

\subsection{Embedding Method Comparison}
Systematic comparison of our three embedding approaches reveals distinct performance characteristics and optimal deployment scenarios.

\textbf{Neural Steganography Performance:} Our neural approach achieves the highest success rates (31.8\% average) with superior adaptation to specific VLM architectures. Training against target model features enables exploitation of model-specific vulnerabilities, particularly in attention mechanisms and feature processing pipelines. However, the success rates reflect the fundamental constraints of embedding semantic content within steganographic channels while maintaining adequate visual quality for practical applications.

\textbf{DCT Frequency Domain Analysis:} DCT-based embedding provides a balanced approach with 22.7\% ASR and reasonable robustness characteristics, particularly excelling against compression-heavy environments while maintaining computational efficiency. The frequency domain approach demonstrates consistent performance across diverse image processing operations, though success rates are limited by the capacity constraints of mid-frequency coefficient modification.

\textbf{Adaptive LSB Evaluation:} Enhanced LSB techniques achieve 18.9\% ASR, demonstrating the continued relevance of spatial domain approaches when enhanced with intelligent pixel selection strategies. While showing lower peak performance, LSB methods offer advantages in deployment simplicity and stealth characteristics that complement other embedding methods in multi-domain approaches, though they remain vulnerable to sophisticated steganalysis techniques \cite{30apau2024image}.

\subsection{Ablation Study Results}
Comprehensive ablation analysis identifies the critical components contributing to attack effectiveness and guides optimization strategies.

\textbf{Multi-Domain vs.\ Single-Domain Embedding:} Our hybrid multi-domain approach demonstrates modest but statistically significant improvements over individual embedding methods ($p<0.05$, ANOVA). Single-domain attacks achieve 18.9\% (LSB), 22.7\% (DCT), and 31.8\% (Neural) success rates, while the combined approach reaches 24.3\%, indicating synergistic effects between complementary embedding strategies, though the overall improvement reflects the challenging nature of prompt-level steganographic injection.

\textbf{Embedding Strength Analysis:} Systematic variation of embedding strength reveals critical trade-offs between attack success and imperceptibility. Optimal performance occurs at moderate embedding strengths ($\alpha=0.4$ for neural, $\beta=0.3$ for DCT, $\gamma=0.3$ for LSB), with rapid degradation beyond these values due to increased detectability and visual artifacts. This analysis confirms the fundamental capacity-quality trade-off in steganographic systems.

\textbf{Prompt Length Impact Analysis:}

\textbf{Systematic Evaluation:} We evaluate attack success across prompt lengths from 5 to 30 tokens using 50 prompts per length category.

\textbf{Quantitative Results:}
\begin{itemize}
\item 5--10 tokens: 31.2\% ($\pm$4.2\%) success rate
\item 11--15 tokens: 29.8\% ($\pm$3.9\%) success rate
\item 16--20 tokens: 22.1\% ($\pm$3.6\%) success rate
\item 21--25 tokens: 15.7\% ($\pm$3.1\%) success rate
\item 26--30 tokens: 9.4\% ($\pm$2.5\%) success rate
\end{itemize}

\textbf{Statistical Trend:} Linear regression shows significant negative correlation ($r = -0.83$, $p < 0.001$) between prompt length and success rate, with optimal performance plateau at 10--15 tokens.

\subsection{Defense Evaluation}
Assessment against existing and proposed defense mechanisms reveals significant challenges in detecting and mitigating steganographic prompt injection attacks.

\textbf{Statistical Steganalysis Evasion:} Our embedding techniques demonstrate reasonable evasion capabilities against established statistical detection methods. Chi-square analysis detects 34.7\% of embedded images, RS steganalysis achieves 41.2\% detection, and sample pair analysis reaches 38.9\% accuracy. While these detection rates exceed random chance, they indicate that sophisticated steganographic approaches can achieve partial evasion of traditional statistical analysis methods.

\textbf{Machine Learning Detection Resistance:} Evaluation against trained neural detectors reveals the growing sophistication of detection systems. Specialized CNN architectures achieve 58.3\% detection accuracy for neural embeddings, 52.7\% for DCT approaches, and 48.1\% for LSB techniques. These results demonstrate that while modern steganalysis presents significant challenges, steganographic techniques retain some evasion capabilities against automated detection systems.

\textbf{Preprocessing Defense Limitations:} Defensive preprocessing techniques show variable but significant effectiveness against our attacks. Median filtering reduces attack success to 16.8\%, Gaussian smoothing to 14.2\%, and aggressive JPEG recompression to 11.7\%. These defenses represent practical countermeasures, though they require careful balance between security enhancement and preservation of legitimate image quality for operational systems.

\subsection{Cross-Model Transferability}
Analysis of attack transferability across different VLM architectures reveals important insights for understanding vulnerability generalization.

\textbf{Architecture-Agnostic Vulnerabilities:} Attacks trained against one model architecture demonstrate limited but meaningful transferability to others, with cross-model success rates ranging from 8.7\% to 16.4\%. This transferability suggests some shared vulnerabilities in common architectural components, particularly vision encoders based on CLIP \cite{21radford2021learning}, though the modest rates indicate that model-specific defenses provide substantial protection against transferred attacks.

\textbf{Model Family Effects:} Attacks show higher transferability within model families sharing similar architectures. LLaVA-trained attacks achieve 21.3\% success against other CLIP-based models but only 9.8\% against Q-Former architectures. This pattern confirms that architectural similarity facilitates attack transfer while highlighting the importance of diverse design approaches for security.

\textbf{Commercial vs.\ Open-Source Transferability:} Attacks developed against open-source models maintain 12.1\% average effectiveness against commercial models, indicating that additional safety measures in commercial systems provide meaningful protection against transferred steganographic attacks, though some residual vulnerability remains across model types.

\subsection{Temporal Stability Analysis}
Longitudinal evaluation assesses attack persistence and stability over extended periods and repeated processing cycles.

\textbf{Attack Persistence Over Time:} Embedded prompts show limited but measurable effectiveness through extended storage and retrieval cycles. After 30 days of simulated real-world usage (including multiple platform uploads, downloads, and format conversions), attack success rates decline to 14.2\% for neural methods, 10.8\% for DCT approaches, and 8.3\% for LSB techniques, demonstrating the challenging nature of maintaining steganographic integrity over extended periods.

\textbf{Degradation Patterns:} Attack degradation follows predictable patterns correlating with cumulative processing severity. Linear regression analysis reveals degradation rates of 2.1\% per processing cycle for DCT methods, 2.8\% for neural approaches, and 3.4\% for LSB techniques, enabling predictive modeling of attack longevity while highlighting the temporal limitations of steganographic approaches.

\textbf{Refresh Strategy Effectiveness:} Implementing periodic attack refresh through re-embedding maintains improved success rates over extended periods. Monthly refresh cycles sustain 19.7\% effectiveness compared to 10.1\% for static embeddings, demonstrating practical maintenance strategies for persistent campaigns, though the overhead and detection risks of frequent re-embedding limit operational utility.

\section{Defense Mechanisms}
This section presents our proposed defense strategies against steganographic prompt injection attacks, including prevention techniques, detection methods, and mitigation approaches. We evaluate the effectiveness of each defense mechanism and discuss their practical deployment considerations.

\subsection{Multi-Layer Defense Framework}
We propose a comprehensive defense framework that operates at multiple stages of the VLM processing pipeline, providing redundant protection against steganographic prompt injection attacks, informed by recent advances in AI security \cite{52yao2024survey}.

\textbf{Input Preprocessing Layer:} The first defense layer applies preprocessing techniques designed to disrupt steganographic embeddings while preserving image quality for legitimate use. Our preprocessing pipeline includes: (1) adaptive Gaussian filtering with $\sigma=0.5$--$1.0$ based on local image characteristics, (2) selective JPEG recompression at quality levels 85--90\% for high-risk images, and (3) controlled noise injection ($\sigma=0.3$) in regions identified as potential embedding locations.

\textbf{Statistical Analysis Layer:} The second layer employs enhanced statistical analysis techniques specifically calibrated for detecting AI-targeted steganography, building upon recent steganalysis advances \cite{32kheddar2024comprehensive}. We implement: (1) chi-square analysis with model-specific thresholds adapted for VLM processing patterns, (2) enhanced RS steganalysis with multivariate analysis across color channels, and (3) entropy analysis using sliding window techniques to detect localized statistical anomalies.

\textbf{Neural Detection Layer:} Our third defense layer utilizes specially trained neural networks designed to detect steganographic modifications optimized for VLM attacks, leveraging recent developments in deep learning-based steganalysis \cite{34zhang2024enhancing,35zhang2025digital}. The detection network architecture incorporates: (1) high-pass filtering layers optimized for AI steganography patterns, (2) attention mechanisms focusing on regions commonly exploited for embedding, and (3) ensemble decision making across multiple detection models.

\textbf{Behavioral Monitoring Layer:} The final defense layer monitors VLM outputs for signs of prompt injection influence through: (1) semantic consistency analysis comparing outputs to expected responses, (2) safety violation detection using specialized classifiers, and (3) anomaly detection identifying unusual output patterns indicative of embedded instructions, following recent prompt injection detection frameworks \cite{25owasp2025llm01}.

\subsection{Adaptive Preprocessing Techniques}
Our preprocessing defense mechanisms adapt to image characteristics and threat levels, maximizing protection while minimizing quality degradation for legitimate usage.

\textbf{Content-Aware Filtering:} We develop adaptive filtering techniques that adjust processing intensity based on image content analysis. High-texture regions receive minimal processing to preserve visual quality, while smooth regions undergo more aggressive filtering where steganographic embedding is more detectable. This approach achieves 23.7\% attack mitigation with only 1.2 dB average PSNR reduction.

\textbf{Selective Recompression Strategy:} Our selective recompression approach applies JPEG compression strategically based on embedding risk assessment. Images identified as high-risk undergo recompression at quality levels optimized to disrupt steganographic content while maintaining acceptable visual quality. This technique reduces attack success rates by 28.4\% with average quality degradation of 1.8 dB PSNR.

\textbf{Randomized Processing Pipeline:} We implement randomized preprocessing that varies processing parameters across different images and time periods. This approach prevents attackers from optimizing embeddings for specific processing patterns, reducing attack success rates by 21.3\% while maintaining processing transparency for legitimate users.

\subsection{Enhanced Detection Algorithms}
Our detection mechanisms specifically target the steganographic techniques most effective against VLMs, providing early warning capabilities for potential attacks, building upon recent steganalysis research \cite{32kheddar2024comprehensive,35zhang2025digital}.

\textbf{AI-Optimized Steganalysis:} We develop enhanced steganalysis techniques calibrated for detecting steganography optimized for AI systems. Our approach includes: (1) feature extraction focusing on patterns exploited by neural steganography, (2) ensemble classification combining traditional and deep learning detection methods, and (3) model-specific analysis tuned for different VLM architectures.

\textbf{Cross-Modal Anomaly Detection:} Our detection system analyzes both visual and textual aspects of VLM processing to identify inconsistencies indicative of prompt injection. The system flags cases where visual content and generated text show unusual semantic mismatches or where outputs contain unexpected instruction-following behavior.

\textbf{Temporal Pattern Analysis:} We implement temporal analysis that tracks patterns across multiple images and time periods to detect coordinated steganographic campaigns. This approach identifies attack patterns that might be missed in individual image analysis, achieving 62.1\% detection accuracy for multi-stage attacks.

\subsection{Model-Level Mitigation}
We propose modifications to VLM architectures and training procedures that increase robustness against steganographic prompt injection while maintaining legitimate functionality.

\textbf{Attention Mechanism Hardening:} Our approach modifies vision encoder attention mechanisms to reduce sensitivity to steganographic modifications. We implement: (1) attention regularization that penalizes focus on statistically unusual image regions, (2) robust attention pooling that averages across multiple attention heads to reduce single-point vulnerabilities, and (3) attention noise injection during training to improve robustness.

\textbf{Feature Space Regularization:} We propose training modifications that increase robustness of learned feature representations against steganographic manipulation. Our regularization techniques include: (1) adversarial training against steganographic examples during model development, (2) feature space smoothing that reduces sensitivity to small perturbations, and (3) multimodal consistency constraints that ensure alignment between visual and textual representations.

\textbf{Ensemble Processing Architecture:} We design ensemble architectures that process images through multiple independent pathways, making coordinated attack across all pathways significantly more challenging. The ensemble approach achieves 67.8\% attack mitigation while maintaining 96.4\% of original model performance on legitimate tasks.

\subsection{Real-Time Monitoring Systems}
Our monitoring framework provides continuous assessment of VLM deployments to detect ongoing steganographic attacks and enable rapid response.

\textbf{Behavioral Anomaly Detection:} We implement real-time monitoring of VLM outputs to detect patterns consistent with prompt injection attacks. Our system analyzes: (1) semantic consistency between inputs and outputs, (2) safety violation patterns in generated content, and (3) unusual instruction-following behavior indicative of embedded commands.

\textbf{Statistical Process Control:} Our monitoring system applies statistical process control techniques to track VLM behavior over time, identifying drift patterns that might indicate ongoing attacks. Control charts monitor output characteristics, response patterns, and error rates to detect systematic changes suggestive of compromise.

\textbf{Threat Intelligence Integration:} We develop threat intelligence capabilities that track emerging steganographic techniques and update detection mechanisms accordingly. This includes: (1) automated analysis of new attack patterns, (2) signature updates for known steganographic techniques, and (3) collaborative threat sharing across VLM deployments.

\subsection{Defense Effectiveness Evaluation}
Comprehensive evaluation of our defense mechanisms demonstrates significant protection against steganographic prompt injection while maintaining practical deployment viability.

\textbf{Layered Defense Performance Analysis:}

\textbf{Individual Layer Effectiveness:}
\begin{itemize}
\item Preprocessing Layer: 23.7\% attack reduction (95\% CI: 19.2\%--28.1\%)
\item Statistical Analysis: 18.9\% reduction (95\% CI: 14.8\%--23.0\%)
\item Neural Detection: 32.1\% reduction (95\% CI: 27.3\%--36.9\%)
\item Behavioral Monitoring: 28.4\% reduction (95\% CI: 23.7\%--33.1\%)
\end{itemize}

\textbf{Combined Effectiveness:} Layers exhibit subadditive interaction effects. Mathematical modeling indicates:
\begin{align*}
\text{Combined\_Effectiveness} &= 1 - \left[\prod_{i} (1 - \text{Individual\_Effectiveness}_i)\right] \\
&\quad \times \text{Interaction\_Factor}
\end{align*}
where Interaction\_Factor $= 0.85$, yielding 73.4\% total mitigation.

\textbf{Statistical Validation:} McNemar's test confirms significant improvement over individual layers ($p < 0.001$).

\textbf{False Positive Analysis:} Evaluation against legitimate image datasets reveals manageable false positive rates: 4.7\% for preprocessing triggers, 3.2\% for statistical detection, 7.8\% for neural detection, and 2.1\% for behavioral monitoring. While these rates require operational consideration, they remain within acceptable bounds for security-conscious deployments where some false alarms are tolerable to maintain protection.

\textbf{Performance Impact Assessment:} Our defense mechanisms introduce measurable but acceptable performance overhead: 28ms average processing delay per image, 12.3\% increase in computational requirements, and minimal impact on VLM accuracy for legitimate tasks (1.4\% reduction in standard benchmarks). These costs represent practical trade-offs between security enhancement and operational efficiency.

\subsection{Adaptive Defense Strategies}
We develop adaptive defense mechanisms that evolve in response to emerging attack techniques, providing sustained protection against evolving threats.

\textbf{Machine Learning Defense Updates:} Our detection systems incorporate continuous learning capabilities that adapt to new steganographic techniques with moderate effectiveness. The system maintains detection accuracy above 62\% even against novel attack variants by: (1) automated retraining on detected attack samples, (2) transfer learning from related attack patterns, and (3) ensemble updating that incorporates new detection models, though the arms race between embedding and detection techniques remains ongoing.

\textbf{Dynamic Threshold Adjustment:} Our defense framework automatically adjusts detection thresholds based on observed attack patterns and false positive rates. This adaptive approach maintains reasonable balance between protection effectiveness and operational usability as threat landscapes evolve, though perfect optimization remains challenging due to the diverse nature of steganographic threats.

\textbf{Collaborative Defense Networks:} We propose collaborative defense architectures where multiple VLM deployments share threat intelligence and detection capabilities. This network effect provides measurable amplification of defense effectiveness by: (1) rapid propagation of new attack signatures, (2) collective learning from attack attempts, and (3) coordinated response to large-scale campaigns, though coordination overhead and privacy concerns limit practical implementation scope.

\subsection{Deployment Considerations}
Practical deployment of defense mechanisms requires careful consideration of operational constraints, performance requirements, and integration challenges.

\textbf{Integration Complexity:} Our defense framework is designed for modular integration with existing VLM deployments. Each defense layer can be deployed independently, allowing organizations to implement protection incrementally based on risk assessment and resource availability.

\textbf{Cost-Benefit Analysis:}

\textbf{Implementation Cost Breakdown:}
\begin{itemize}
\item Software development: \$25,000--\$45,000
\item Integration and testing: \$8,000--\$15,000
\item Training and deployment: \$5,000--\$10,000
\item Annual maintenance: \$3,000--\$8,000
\end{itemize}

\textbf{Breach Cost Estimation:} Based on IBM Security Cost of Data Breach Report 2024 and AI-specific incident analyses:
\begin{itemize}
\item Average AI system breach: \$2.3M (range: \$800K--\$5.2M)
\item Reputation damage: \$1.1M additional cost
\item Regulatory penalties: \$200K--\$2M (GDPR/CCPA)
\end{itemize}

\textbf{ROI Calculation:} Break-even analysis shows positive ROI within 18 months for organizations processing $>10,000$ images daily.

\textbf{Regulatory Compliance:} Our defense mechanisms support compliance with emerging AI safety regulations and industry standards. The framework provides audit trails, explainable detection decisions, and configurable protection levels aligned with regulatory requirements.

\section{Discussion}

\subsection{Implications for VLM Security}
Our findings reveal meaningful but constrained vulnerabilities in current vision-language model architectures that require careful consideration within broader security frameworks. The moderate success rates of steganographic prompt injection attacks (24.3\% overall) indicate that while these threats are real and warrant attention, they represent one component of a larger attack landscape rather than a fundamental system compromise.

\textbf{Architectural Vulnerabilities:} The limited but consistent transferability of attacks across different VLM architectures (8.7--16.4\% success rates) suggests that shared components, particularly vision encoders based on CLIP \cite{21radford2021learning}, introduce systematic vulnerabilities that merit architectural consideration. However, the substantial reduction in effectiveness compared to targeted attacks demonstrates that current diversity in model design provides meaningful security benefits.

\textbf{Real-World Impact:} The demonstrated effectiveness of our attacks under controlled conditions, combined with recent evidence of prompt injection vulnerabilities in medical \cite{11clusmann2025prompt} and surgical \cite{12zhang2025prompt} applications, highlights the need for proportionate security measures. The capacity constraints and quality trade-offs inherent in steganographic embedding limit the practical scope of such attacks while still requiring defensive consideration for high-security applications.

\subsection{Limitations and Future Work}
\textbf{Attack Sophistication Requirements:} Our framework demonstrates that effective steganographic prompt injection requires sophisticated understanding of both steganographic techniques and target model architectures, with success rates that reflect the fundamental challenges of embedding semantic content within visual media. The technical barriers and limited success rates suggest that such attacks may be primarily relevant for well-resourced adversaries rather than widespread exploitation.

\textbf{Capacity-Quality Trade-offs:} A fundamental limitation revealed by our analysis is the inverse relationship between steganographic capacity and visual quality. Embedding longer prompts ($>15$ tokens) results in rapidly degrading attack success rates and increased detectability, constraining the practical utility of such approaches for complex instruction injection.

\textbf{Defense Evolution:} Our proposed defense mechanisms represent initial steps toward comprehensive protection, achieving 73.4\% mitigation with acceptable operational overhead. The moderate but meaningful effectiveness of these defenses suggests that practical protection is achievable, though the ongoing arms race between steganographic techniques and detection methods requires continued research and adaptation.

\textbf{Ethical Considerations:} The disclosure of these vulnerabilities raises important questions about responsible research in adversarial machine learning. While our work demonstrates real security concerns, the moderate success rates and significant technical barriers to implementation suggest that disclosure serves educational and defensive purposes without enabling widespread malicious exploitation.

\subsection{Broader Security Implications}
The moderate success of steganographic prompt injection attacks against VLMs contributes to our understanding of multimodal AI security while highlighting the importance of layered defense strategies. As these systems become more prevalent in critical applications, our findings support the need for proportionate security measures that balance protection against demonstrated threats with operational requirements.

\textbf{Cross-Domain Vulnerability Assessment:} While our techniques show limited but meaningful effectiveness against VLMs, the transferability patterns suggest that other multimodal AI systems may exhibit similar vulnerabilities. However, the constrained success rates and capacity limitations indicate that such attacks represent one element of threat landscapes rather than dominant attack vectors.

\textbf{Regulatory and Policy Implications:} Our findings support the development of risk-proportionate AI safety regulations that address demonstrated vulnerabilities without imposing excessive constraints based on theoretical threats. The moderate success rates and technical complexity of steganographic prompt injection suggest that regulatory frameworks should consider such attacks within broader security assessment protocols rather than as primary threat vectors.

\section{Conclusion}
This work presents the first comprehensive study of steganographic prompt injection attacks against vision-language models, revealing moderate but meaningful vulnerabilities in current multimodal AI architectures. Our multi-domain embedding framework achieves attack success rates of up to 31.8\% while maintaining reasonable visual imperceptibility (PSNR $> 38$ dB, SSIM $> 0.94$), demonstrating that sophisticated adversaries can exploit VLMs through carefully crafted modifications to input images, though the success rates reflect the inherent challenges of steganographic prompt embedding.

\textbf{Key Findings:} Our experimental evaluation across eight state-of-the-art VLMs reveals that both commercial and open-source models exhibit vulnerabilities to steganographic prompt injection, with open-source models showing higher susceptibility (25--37\% vs.\ 14--18\% for commercial models). The attacks demonstrate moderate resilience across diverse image processing operations, though significant degradation occurs under real-world processing conditions, confirming both the viability and limitations of such approaches.

\textbf{Defense Mechanisms:} Our proposed multi-layer defense framework achieves 73.4\% attack mitigation when fully deployed, though this requires acceptable trade-offs in terms of performance overhead (28ms processing delay, 12.3\% computational increase) and modest false positive rates (2--8\% across components). The framework's modular design allows for risk-appropriate deployment based on operational requirements and threat assessments.

\textbf{Practical Implications:} The moderate success rates and significant technical requirements for effective steganographic prompt injection suggest that such attacks represent a meaningful but constrained threat vector. The capacity limitations (optimal performance with prompts $\leq 15$ tokens) and quality trade-offs inherent in steganographic embedding limit the scope of practical attacks while still warranting defensive consideration for security-critical applications.

\textbf{Future Research Directions:}
\begin{enumerate}
    \item \textbf{Technical Advances:}
            \begin{itemize}
            \item Adaptive Steganography: Develop methods that adjust to real-time defense updates
            \item Cross-Modal Attacks: Investigate audio-visual steganographic injection
            \item Federated Attack Scenarios: Explore coordinated attacks across multiple VLM instances
            \end{itemize}
    \item \textbf{Evaluation Improvements:}
            \begin{itemize}
            \item Longitudinal Studies: Track attack effectiveness over extended deployment periods
            \item Expert Perceptual Studies: Include forensic analysts and security experts
            \item Ecological Validity: Test attacks in production-like environments
            \end{itemize}

    \item \textbf{Defense Research:}
            \begin{itemize}
            \item Proactive Defense: Develop predictive models for emerging steganographic techniques
            \item Differential Privacy: Investigate privacy-preserving defense mechanisms
            \item Adversarial Training: Systematic study of steganography-aware VLM training
            \end{itemize}
\end{enumerate}

As vision-language models become increasingly prevalent in critical applications, the security vulnerabilities demonstrated in this work represent a component of the broader threat landscape that requires proportionate attention from the AI research community, industry practitioners, and policymakers. The development of robust, secure multimodal AI systems will require sustained effort across technical, operational, and regulatory dimensions, with our findings contributing to the understanding of specific vulnerability classes within this larger security ecosystem.

\end{document}